\documentclass[amsmath, superscriptaddress, amssymb, aps, prd,nofootinbib,onecolumn,10pt]{revtex4-1}

\usepackage{float}
\usepackage{epsfig}
\usepackage{dsfont}
\usepackage{amsmath,amssymb}
\usepackage{braket}
\usepackage{amsfonts}
\usepackage{mathtools}
\usepackage{graphicx}  
\usepackage[dvipsnames]{xcolor}
\usepackage{todonotes}
\usepackage{subcaption}
\usepackage{bbm}
\usepackage{tikz}
\usetikzlibrary{arrows,decorations.markings}
\usepackage{multirow}
\usepackage{lmodern}
\usepackage{braket}
\usepackage{wrapfig}
\usepackage{comment}
\usepackage{xcolor}
\RequirePackage[colorlinks,citecolor=blue,urlcolor=magenta,linkcolor=blue]{hyperref}
\usepackage[english]{babel}
\usepackage{esint}
\usepackage{dsfont}
\usepackage{soul}
\usetikzlibrary{decorations.pathmorphing}
\tikzset{snake it/.style={decorate, decoration=snake}} 
\usetikzlibrary{shadings} 

\newcommand{\lsim}{\raisebox{-0.13cm}{~\shortstack{$<$ \\[-0.07cm]
      $\sim$}}~}

\begin{document}

\title{Are accelerated detectors sensitive to Planck scale changes?}

\author{Harkirat Singh Sahota}
\affiliation{Department of Physical Sciences, Indian Institute of Science Education \& Research (IISER) Mohali, Sector 81 SAS Nagar, Manauli PO 140306 Punjab India.}
\affiliation{Department of Physics,  Indian Institute of Technology Delhi, Hauz Khas, New Delhi, 110016, India.}
\email{harkirat221@gmail.com, ph17078@iisermohali.ac.in}
\author{Kinjalk Lochan}
\email{kinjalk@iisermohali.ac.in}
\affiliation{Department of Physical Sciences, Indian Institute of Science Education \& Research (IISER) Mohali, Sector 81 SAS Nagar, Manauli PO 140306 Punjab India.}

\begin{abstract}
One of the foremost concerns in the analysis of quantum gravity is whether the locations of classical horizons are stable under a full quantum analysis. In principle, any classical description, when interpolated to the microscopic level, can become prone to fluctuations. The curious question in that case is if there indeed are such fluctuations at the Planck scale, do they have any significance for physics taking place at scales much away from the Planck scale? In this work, we address the question of small scales and address whether there are definitive signatures of Planck scale shifts in the horizon structure. In a recent work \cite{Lochan:2021pio}, it was suggested that in a nested sequence of Rindler causal wedges, the vacua of preceding Rindler frames appear thermally populated to a shifted Rindler frame. The Bogoliubov analysis relies on the global notion of the quantum field theory and turns out to be insensitive to the local character of these horizon shifts. We investigate this system by means of the Unruh-DeWitt detector and see if this local probe of the quantum field theory is sensitive enough to the shift parameters to reveal any microscopic effects. For the case of infinite-time response, we recover the thermal spectrum, thus reaffirming that the infinite-time response probes the global properties of the field. On the other hand, the finite-time response turns out to be sensitive to the shift parameter in a peculiar way that any detector with energy gap $\Omega c/a \sim 1$ and is operational for timescale $T a/c \sim 1$ has a measurably different response for a macroscopic and microscopic shift of the horizon, giving us a direct probe to the tiniest separation between the Rindler wedges. Thus, this study provides an operational method to identify Planck scale effects that can be generalized to various other interesting gravitational settings. 
    
\end{abstract}
\maketitle

\section{Introduction}
    
    In the discourse of quantum gravity, it is usually accepted that the spacetime described by a smooth metric is only a macroscopic effective description arising out of some microscopic degrees of freedom. Perhaps at the microscopic level (traditionally believed to be the Planck scale), the degrees of freedom describing spacetime geometry would be quantum in nature. Their basic quantum character would, therefore, be most visible at the Planck scale itself \cite{Percival:1995an}, where the quantum uncertainty is expected to impart quantum fluctuations to the geometric description. 
    
    Even much before the Planck scale, at the semiclassical level itself, since the dynamics of spacetime is essentially dictated by matter, which is quantum in nature, it is also possible that the quantum matter imparts some uncertainty to the gravitational sector \cite{Hu:2003qn}. Thus, there may be induced fluctuations \cite{Slacker:1958} from the matter side apart from any potential inherent uncertainty quantum gravity offers. Therefore, the macroscopic spacetime structure or the symmetries of the metric that we hold dear might not hold in their truest form as we march towards probing smaller and smaller length scales. This may lead to distortion in the ``classical metric" based analysis of modes that probe such scales, such as the dispersion relation of a free quantum field.  It is worth considering whether there can be gedanken experiments that can capture and relay such effects to the macroscopic scales \cite{Husain:2015tna}. 
   
    In that spirit, it is interesting to consider if light cones and various horizons are stable classical objects or if they also have some intrinsic quantum character, which macroscopic observations are oblivious about \cite{Ford:1994cr}. This particular thought becomes more pertinent in the context of black hole kinematics as well as its dynamics \cite{Ford:1997zb,Thompson:2008vi}. If horizons do have some intrinsic quantum character, then black hole horizons may also be fuzzy \footnote{There are proposals suggesting that black holes being high entropy objects catapult their microscopic fluctuations even to their macroscopic horizons \cite{Mathur:2005zp}.} and the standard analysis relying upon the availability of a definitive horizon (such as the Hawking-Unruh effect) may have some bearing \cite{Ford:1997zb}.  

\begin{figure}[ht]
    \begin{subfigure}{.49\textwidth}
    \centering
    \begin{tikzpicture}[scale=1.3]
            \draw[orange, snake it] (0,1.5) -- (1.5,1.5);
            \draw[black,bend right = 10] (0,-3) to (0.4,1.5);
            \draw[black, dotted, bend right = 20] (0,-3) to (1.5,1.5);
            \draw[blue, dotted, bend right = 15] (0,-3) to (1.3,1.5);
            \draw[black] (0,-3) -- (3,0) -- (1.5,1.5);
            \draw[black] (0,-3) -- (0,1.5);
            \draw[blue] (1.6,-1.4) -- (0,0.2) -- (0.7,0.9); 
            \draw[blue, dashed] (0.7,0.9) -- (1.3,1.5);
            \draw[green, bend right = 10] (0.65,0.83) -- (0.85,0.85);
            \draw[green, thick] (0.85,0.85) -- (1.5,1.5);
            \draw[red, thick] (2.2,-0.8) -- (0,1.5);
            \draw[red, thick] (2.3,-0.7) -- (0.2,1.5);
            \node[label=right:$\mathcal{J}^+$] (4) at (2.3,0.8) {};
            \node[label=right:$\mathcal{J}^-$] (5) at (1.5,-2) {};
            \node[label=above: $i^{+}$] at (1.5,1.5) {};
            \node[label=below: $i^{-}$] at (0,-3) {};
            \node[label=right: $i^{0}$] at (3,0) {};
            \fill[blue!50,nearly transparent]  (0,-3) -- (0,1.5) decorate[decoration=snake] {  -- (0.4,1.5)  } to[bend left=10] cycle;
            \fill[red!50,nearly transparent]  (2.2,-0.8) -- (0,1.5) -- (0.2,1.5) -- (2.3,-0.7) -- cycle;
    \end{tikzpicture}
                    \caption{Black hole horizon shifts with the addition of mass. The exterior observers have to adjust their accelerations to remain outside, lest they fall in}\label{Penrose}
\end{subfigure}
\begin{subfigure}{.5\textwidth}
    \centering
     \includegraphics[scale=0.95]{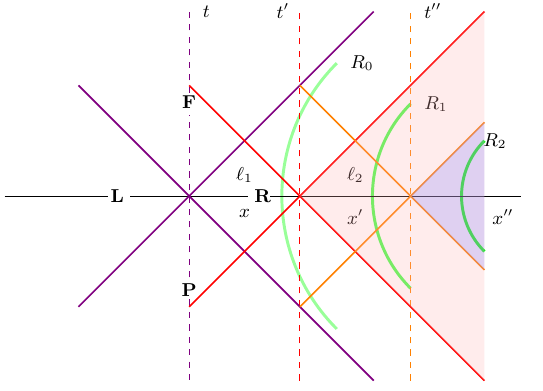}
        \caption{Nested Rindler structure: Accelerated observers on Rindler trajectories find the vacuum state of all preceding Rindler observers thermally populated \cite{Lochan:2021pio}}\label{NestedRindler}
    \end{subfigure}
    \caption{Shifts in the horizon structure. In the first case, it is achieved by the addition of an infinitesimal mass in the black hole and a microscopic shift in the Rindler wedges in the second case.}
\end{figure}
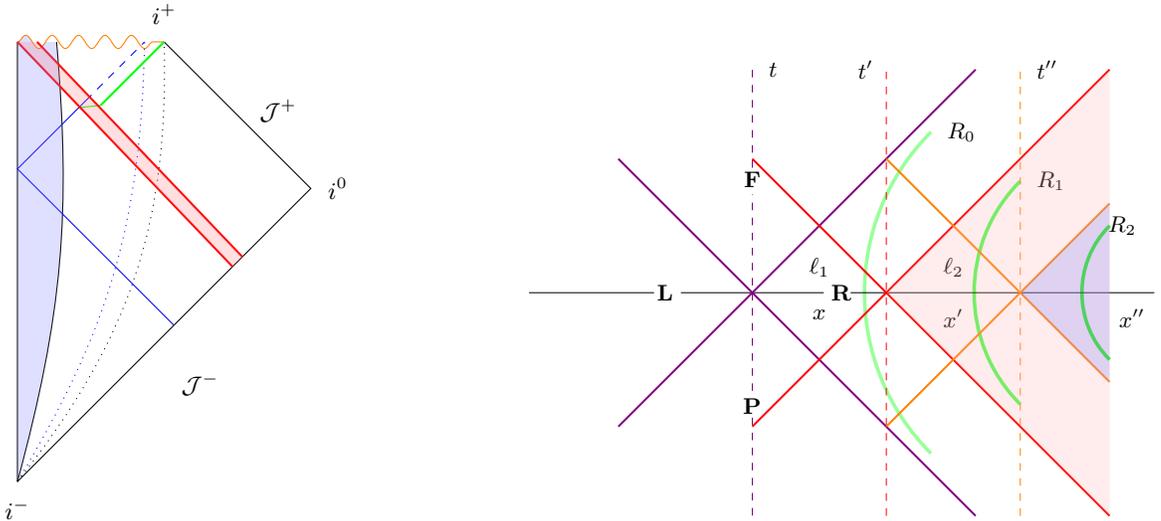

   
    Thus, the question of real relevance is if there are macroscopic detectors that can decisively capture Planck scale effects i.e., if the black hole horizons shift by Planck scale during their formation (or evaporation), do they leave any measurable imprint of their shift to the outside geometry at the macroscopic level? 
    Such effects, if present, can have important implications for the black hole information paradox as well \cite{Chakraborty:2017pmn}.  The accelerated trajectories that remain outside the black hole and are well placed to judge the shift in the horizon, shown by the dotted curves in Fig. (\ref{Penrose}). Thus, it is more precisely a problem if detectors on accelerated trajectories can sense Planck scale shifts in a finite time duration of operation.
    
    In an attempt of answering these questions, we pose the problem in the Rindler world in this work. Since in the Rindler space, there is truly no dynamical character to the horizon (at least classically), we investigate if two Rindler observers which have different horizons can distinguish each other through their detectors' responses, particularly in the case when the horizons are microscopically separated (say by Planck scale shift). We investigate this system using a local probe of the field content, a finite-time Unruh-DeWitt (UDW) detector response \cite{Unruh:1976db}. The UDW detector is an operational quantum device that probes the quantum field configurations and is used to study the quantum effects in a noninertial setting or gravitational setting \cite{Crispino:2007eb,Louko:2007mu}. 
    There are indications in the literature that its low-energy response may be sensitive to Planck-scale physics -- say, potential Lorentz violations at the Planck scale -- implemented via a modified dispersion relation \cite{Husain:2015tna} and a polymerized scalar field \cite{Kajuri:2015oza}.


    As suggested in \cite{Lochan:2021pio}, in a nested system of Rindler observers (see Fig. (\ref{NestedRindler})), the number expectation "turns on"\footnote{For no shift between the light cones, the number operator has zero expectation value.} to full thermal behavior even for Planck order shift between Rindler horizons and remains insensitive to the amount of the shift thereafter, one needs to check if the UDW detector also responds in a similar fashion, if it has a global dependence (always "on" detector). More interestingly, it will be interesting to learn if that really is the case, and then how much of this feature survives for a finite time operation. Further, if it deviates away from a thermal description for the finite-time response, whether the non-thermal character contains enough information to distinguish a macroscopic shift from a Planck order shift. In this light, we explore detectors in such nested configurations that operate for a finite time and estimate their response {\it vis \'a vis} a macroscopic and microscopic shift.  



\section{Nested frames and thermal response}

We consider a UDW detector on an accelerating trajectory of $R_1$, which is observing the vacuum of $R_0$, the trajectory given by
\begin{align}
    a_0^{-1}e^{a_0\zeta_0}\sinh(a_0\tau_0)=a_1^{-1}e^{a_1\zeta_1}\sinh(a_1\tau_1),\quad a_0^{-1}e^{a_0\zeta_0}\cosh(a_0\tau_0)-\ell_1=a_1^{-1}e^{a_1\zeta_1}\cosh(a_1\tau_1).
\end{align}
Here $(\tau_{i},\zeta_{i})$ is the coordinate chart for the $1+1$ Rindler wedge $R_i$, $a_i$ is the acceleration of the Rindler observer $R_i$, and $\ell_1$ is the separation between the Rindler wedges of $R_0$ and $R_1$, as shown in Fig. (\ref{NestedRindler}). The probability of transition from the ground state $\ket{E_0}$ to the excited state $\ket{E_1}$ of detector is \cite{Birrell}
\begin{align}
	P_{0\rightarrow1}(\Omega)= \bar{m}_{10}^2\int_{-{\cal T}}^{{\cal T}}d\tau_1\int_{-{\cal T}}^{{\cal T}}d\tau_1'e^{-i\Omega(\tau_1-\tau_1')}\chi(\tau_1)\chi(\tau_1')G(\tau_1,\tau'_1) =\bar{m}_{10}^2F(\Omega),\label{TrPr} 
\end{align}
where $\bar{m}_{10}$ is the transition element of the monopole moment $\braket{E_1|\hat{m}(0)|E_0}$, $\Omega=(E_1-E_0)/\hbar$ is the energy gap, $\mathcal{T}$ is the long time operation within the first order perturbation theory approach, $F(\Omega)$ is called the response function, and $\chi(\tau)$ is the window function that models the switching of detection. $G(\tau_1,\tau_1')$ is the pullback of the Green's function in state $\ket{0}_{R_0}$\footnote{The in-state can be prepared in the $\ket{0}_{R_0}$ state with a high fidelity by performing a number operator measurement by $R_0$ over the inertial vacuum state, particularly when the acceleration $a_0$ is small. Though as argued in \cite{Lochan:2021pio}, the Wightmann function has the same Hadamard structure, the leading order result remains the same for any general state in the Rindler Fock basis.} on the detector trajectory \cite{Lochan:2021pio}
\begin{align}
	G(\tau_1,\tau'_1)=\int_0^\infty\frac{d\omega}{2\pi\omega}\left(e^{a_1(\tau_1-\tau'_1)}\right)^{-i\frac{\omega}{a_0}}\left[\frac{1+a_1\ell_1e^{a_1\tau_1}}{1+a_1\ell_1e^{a_1\tau'_1}}\right]^{i\frac{\omega}{a_0}}.
\end{align}
The two-point function is computed for the state $\ket{\psi}=\ket{0}_{R_0}\otimes\ket{0}_{L_0}$ and the field operators $\hat{\phi}$ restricted in the right wedge of the shifted frame is complemented with the identity operator in the left wedge of shifted frame. The response function, in this case, can be cast in the form
\begin{align}
	F(\Omega)&=\int_0^\infty\frac{d\omega}{2\pi\omega}|I(\omega)|^2,\\
	I(\omega)&=\int_{-{\cal T}}^{{\cal T}}d\tau e^{-i\left(\Omega+\omega\frac{a_1}{a_0}\right)\tau}\chi(\tau)\left(1+a_1\ell_1e^{a_1\tau}\right)^{i\frac{\omega}{a_0}}.\label{Io}
\end{align}
For thermal behavior, we are interested in the case when the detector is switched on for a  very long time, i.e., when $\chi(\tau)=1$ along with ${\cal T} ( \gg \Omega^{-1})  \rightarrow \infty$. In this case, the integral approaches  delta function with a positive definite argument in the limit $\ell_1\rightarrow 0$, causing the response function and thus transition probability to vanish. Therefore, in the case of no shift in the horizons, the detector does not click. For finite $\ell_1$, on the other hand, the integral \eqref{Io} in the long time limit  $\mathcal{T}\rightarrow\infty$  (henceforth we work in this limit only) can be expressed as \cite{Oberhettinger1990}
\begin{align}
	I(\omega)=\frac{(a_1\ell_1)^{i\left(\frac{\Omega}{a_1}+\frac{\omega} {a_0}\right)} \Gamma\left(i\frac{\Omega}{a_1}\right)\Gamma\left(-i\left(\frac{\Omega}{a_1}+\frac{\omega} {a_0}\right)\right)}{a_1\Gamma\left(-i\frac{\omega}{a_0}\right)}.\label{IwO}
\end{align}
Using the property of the Gamma functions $\displaystyle \Gamma(i\nu)\Gamma(-i\nu)=\pi/\nu\sinh{\pi\nu}$, the response function can be expressed as
\begin{align}
	F(\Omega)=\frac{1}{a_1\Omega\sinh{\left(\frac{\pi\Omega}{a_1}\right)}}\int_0^\infty\frac{d\omega}{2a_0}\frac{\sinh{\frac{\pi\omega}{a_0}}}{\left(\frac{\Omega}{a_1}+\frac{\omega} {a_0}\right)\sinh{\left(\frac{\pi\Omega}{a_1}+\frac{\pi\omega} {a_0}\right)}}.
\end{align}
Interestingly, this expression matches with the number expectation of $R_1$ in vacuum $\ket{0}_{R_0}$, and the integral is UV divergent with logarithmic divergence, as seen by the asymptotic behavior of the integrand, see Appendix \ref{App2}. The integral can be cast in a form with a divergent contribution along with finite subleading contributions \cite{Lochan:2021pio}
\begin{align}
	F(\Omega)=\frac{1}{a_1\Omega}\left[\frac{\pi\delta(0)}{e^{\frac{2\pi\Omega}{a_1}}-1}-\frac{\log\frac{\Omega}{a_1}}{e^{\frac{2\pi\Omega}{a_1}}-1}-\sum_{n=1}^\infty\Gamma\left(0,2n\frac{\pi\omega}{a_1}\right)\right]=\frac{2\pi}{\Omega} N_\Omega,
\end{align}
where the divergence in the first term indicates the fact that the detector is switched on for a long time at a constant rate. The response rate, in this case, can be obtained by dividing the response function by the {\it switch-on} time\footnote{In this analysis, we use the response (transition) rate defined as response function (transition probability) divided by the detector switch-on time \cite{Unruh:1976db,Svaiter:1992xt}. One can equivalently define response rate as a time derivative of response function\cite{Sriramkumar:1999nw}, and both definitions lead to the same thermal response for the case when the detector is switched on for large time $\Omega T\gg 1$, but differ marginally for $T\lsim\Omega^{-1}$, as shown in Appendix \ref{App3}. Any choice of the detector response rate does not change our results qualitatively.}, formally expressed as $2\pi\delta(0)$ for a detector switched on for a long time\footnote{For translation invariant systems, the response function integral can be written as $\displaystyle\int_{-\infty}^\infty d\tau_+\int_{-\infty}^\infty d\tau_- e^{-i\Omega\tau_-}G(\tau_-)$ and is independent of integration variable $\tau_+$. We obtain the  average response rate for a large switch-on time for the detector by approximating it with Dirac-delta distribution  in frequency space, which gives the total range of switch-on time  $\displaystyle2\pi\delta(0)=\lim_{{\cal T}\gg \Omega^{-1}}\int_{-{\cal T} }^{{\cal T}} d\tau_+$.} 
\begin{align}
	R_{0\rightarrow 1}(\Omega)=\frac{F(\Omega)}{2\pi\delta(0)}=\frac{1}{2a_1\Omega}\frac{1}{e^{\frac{2\pi\Omega}{a_1}}-1}.
\end{align}
This response rate verifies that the infinite time response is indeed thermal for any finite $\ell_1$ (in fact, insensitive to the magnitude of $\ell_1$), as suggested in \cite{Lochan:2021pio}. {\it Therefore, this detector will turn on to a full thermality even for a Planck scale shift between the horizons.}  Thus, the detector responds {\it nonperturbatively} for a microscopic shift, but such a response is not a definitive signature of Planck scale shift as the response has a degeneracy for any magnitude of the shift. Therefore, we now focus on the response of finite-time detector in an attempt to break the degeneracy of scales, motivated by the observation that the shift dependence appears only as a phase in Eq. (\ref{IwO}) for an infinite time response, possibly not for finite-time responses. If the finite time response remains appreciably strong and develops a dependency on the shift parameter in addition to that existing through the phase, then one can hope to have distinguishable responses for different amounts of shifts. We will see next that this precisely is the case.


\section{Finite time response for Nested Rindler detector}
In a realistic physical scenario, any detector will remain switched on only for a finite duration, where ideally, one should introduce a compactly supported window function in the interaction term. With a smooth function $\chi(\tau)$ modeling the switching profile \cite{Satz:2006,Louko:2006zv}, the response function can be obtained from \eqref{TrPr}, with the modification in $I_1(\omega)$ as
\begin{align}
	I_1(\omega)=\int_{-\infty}^\infty d\tau \;e^{-i\left(\Omega+\omega\frac{a_1}{a_0}\right)\tau}e^{-|t|/T}\left(1+a_1\ell_1e^{a_1\tau}\right)^{i\frac{\omega}{a_0}}.\label{io1}
\end{align}
Here the window function with exponential tail above a characteristic timescale $T$  is considered \cite{Sriramkumar:1994pb}, and for the case $a_1\ell_1<1$, we obtain
\begin{align}
	I_1(\omega)=\frac{(a_1\ell_1)^{\frac{1}{a_1T}} \Gamma\left(i\frac{\Omega}{a_1}+\frac{1}{a_1T}\right)\Gamma\left(-i\left(\frac{\Omega}{a_1}+\frac{\omega}{a_0}\right)-\frac{1}{a_1T}\right)}{a_1\Gamma\left(-i\frac{\omega}{a_0}\right)}+\frac{2T}{\Gamma\left(-i\frac{\omega}{a_0}\right)}\sum_{n=0}^{\infty}\frac{(-1)^n(a_1\ell_1)^{-i\left(\frac{\Omega}{a_1}+\frac{\omega} {a_0}\right)+n}\Gamma\left(-i\frac{\omega}{a_0}+n\right)}{\left(1-a_1^2T^2\left(-i\left(\frac{\Omega}{a_1}+\frac{\omega} {a_0}\right)+n\right)^2\right)n!},
\end{align}
(see the derivation in Appendix \ref{App1}).

The first term gives the thermal response in the limit $T\rightarrow\infty$,  while the series vanishes for the infinite time detector, as expected. For finite $T$, the infinite series is the dominant contributor to the response in the limit $\ell_1\rightarrow0$ (interestingly, its expression is exactly the same as a finite-time inertial detector in the inertial vacuum, see Appendix \ref{App12}), as the thermal term drops off for vanishingly small $\ell_1$. It is the finite $T$ and $\ell_1\neq0$ case, that is the most interesting. In this case, the leading order response can succinctly be expressed as\footnote{In the $a_1\ell_1\ll 1$ limit, which we are interested in, the higher order terms in the series are suppressed as they fall off as $(a_1 \ell_1)^{n}$.}

\begin{align}
	R_{0\rightarrow1}(\Omega)=\frac{1}{T}\int_0^\infty\frac{d\omega}{2\pi\omega}\left|\frac{ (a_1\ell_1)^{\frac{1}{a_1T}}\Gamma\left(i\frac{\Omega}{a_1}+\frac{1}{a_1T}\right)\Gamma\left(-i\left(\frac{\Omega}{a_1}+\frac{\omega} {a_0}\right)-\frac{1}{a_1T}\right)}{a_1\Gamma\left(-i\frac{\omega}{a_0}\right)}+\frac{2T(a_1\ell_1)^{-i\left(\frac{\Omega}{a_1}+\frac{\omega} {a_0}\right)}}{\left(1-a_1^2T^2\left(-i\left(\frac{\Omega}{a_1}+\frac{\omega} {a_0}\right)\right)^2\right)}\right|^2.\label{tpft}
\end{align}

For a discussion on the convergence of this integral, see Appendix \ref{App2}. We see that the shift dependence appears as a modulating factor with the (would be) thermal contribution and as a phase factor with the correction to the thermality in terms of an inertial-like contribution. The modulating effect of the shift-dependent term is most pronounced at some finite $T$, and it is gradually washed away as the detector is kept on for a longer duration. One can quickly check from Eq. \eqref{tpft} that the $\ell_1$ dependency is again washed out for $T\rightarrow 0$, as potential thermal term drops off in this limit and $\ell_1$ remains as a pure phase. Thus, $dR_{0\rightarrow1}/d\ell_1\rightarrow 0$ for both $T\rightarrow 0,\;\infty$. However, between these extremes, there is a nontrivial dependence on $\ell_1$, i.e. $dR_{0\rightarrow1}/d\ell_1\neq 0$. Since the $\ell_1$ dependency in the response creeps in whenever $T$ is nonzero and finite, it is instructive to find out for what $T$ the $\ell_1$ dependency is most pronounced and if it is distinguishably apart for a macroscopic shift and a microscopic shift.


\begin{figure}[H]
    \centering
    \begin{subfigure}[b]{0.49\textwidth}
    \centering
    \includegraphics[width=\textwidth]{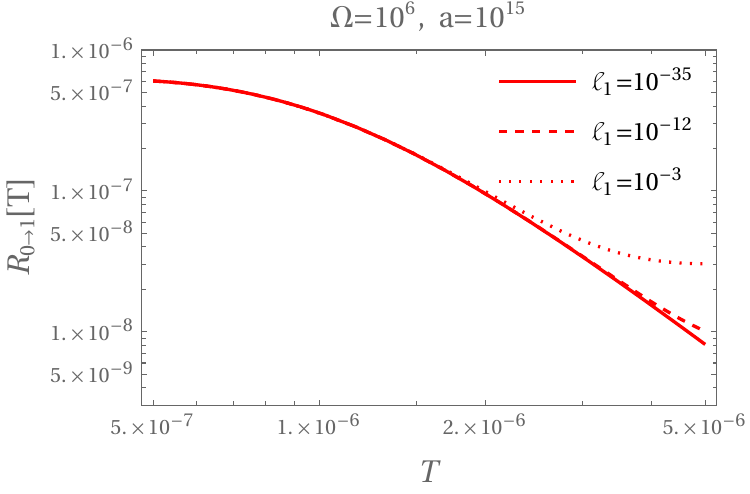}
    \subcaption[]{}
    \label{sf1}
    \end{subfigure}
    \hfill
    \begin{subfigure}[b]{0.49\textwidth}
    \centering
    \includegraphics[width=\textwidth]{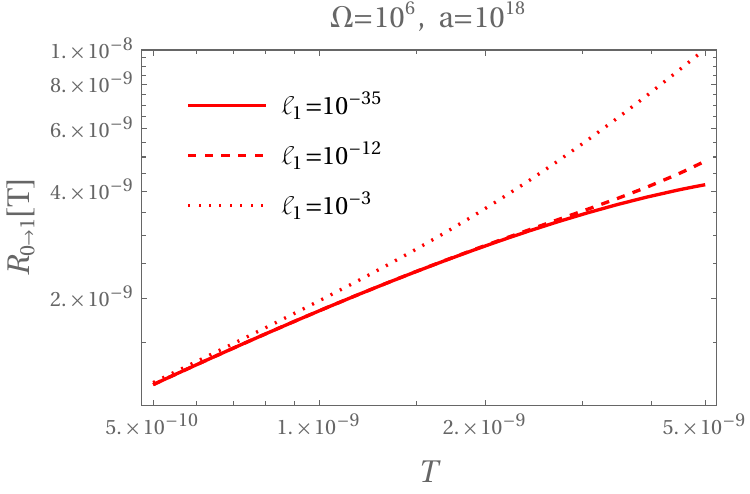}
    \subcaption[]{}
    \label{sf2}
    \end{subfigure}
    \caption{The response rate as a function of the width of the window function for different accelerations in Fig. \ref{sf1} and Fig. \ref{sf2}, for the shifts of Planck length (solid curve) and macroscopic order (dashed curve).}
    \label{resrate}
\end{figure}

If the timescale of detector operation is small enough, then the second term in Eq. \eqref{tpft} dominantly determines the response, as the first term falls rapidly through its dependency on $T$ (essentially as $\sim (a_1\ell_1)^{1/a_1T}$ for small $a_1T$ and $a_1\ell_1<1$), while the fall of the second term is linear with respect to $T$. Yet, if the first term becomes completely irrelevant, then the $\ell_1$ dependency is again erased, as it survives only in the phase of the inertial-like term. Therefore, there should be a window, i.e., $a_1T\sim 1$, where both the first and the second term in Eq.(\ref{tpft}) are of importance for a macroscopic shift (in fact, the first few subleading terms in the series will contribute as well in this case). For the same $T$ if the shift is of the Planck order, then the first term, through its sharp dependence on $\ell_1$ in the modulating factor  $(a_1\ell_1)^{1/a_1T}$, damps down considerably. Thus, for the microscopic shift, only the second term determines the effective response, whereas for the macroscopic shifts, both terms contribute. This, therefore, would lead to a differentiable change in the response of the detector between a large and a small shift, which one can utilize to estimate the difference between the Rindler wedges of two Rindler observers. As discussed in the introduction, a similar computation can be envisaged to estimate the shift of the black hole horizon as seen by two exterior observers before and after the introduction of a mass shell that falls into the black hole. Thus, such a UDW probe can be expected to naturally show up the shift dependency for a finite time operation $1<a_1T <a_1T_{th}$, where $T_{th}$ is the time the detector takes to attain a thermal rate.

To illustrate this point with clarity, we integrate the expression in Eq. \eqref{tpft} numerically after restoring the factors of $c$ and $\hbar$ and selecting the parameters such that the tail contribution from the UV sector in the frequency integral can be safely ignored for a suitable cutoff\footnote{One can numerically verify that the contribution in the tail post the cutoff is vanishingly small and numerical results can be trusted for $T<c/a_1$, the regime of interest, see discussions in Appendix \ref{App2}}, i.e., $c\omega/a_1\rightarrow10^{10}$. To deal with the IR divergence of the inertial response in $1\oplus 1$ dimension \cite{Louko:2014aba}, we also have to employ an IR cutoff. With these choices, we have plotted the response rate as a function of time for a detector of energy gap $\Omega=10^6s^{-1}$ in Fig. \ref{resrate}. 

We see that the modulating effects of $\ell_1$ start appearing if the detector is kept on for a sufficient duration (microseconds for the acceleration of order $10^{15}ms^{-2}$ and nanoseconds for the acceleration of order $10^{18}ms^{-2}$). Though the numerical accuracy considerations compel us to keep the parameter $T<c/a_1$, the trends indicate that the difference in the transition rate for the Planck length shifts and macroscopic shifts keeps on increasing as the detector is kept on for a longer duration, along the expected lines. We anticipate that the difference will reach the maximum for a finite $T$, and the two expressions will merge onto the thermal response for a very large $T$. However, the numerical techniques employed here keep the integral trustworthy for a finite duration, as used in Fig. \ref{resrate}, which prevents us from locating the extrema. However, even within this allowed range, the response becomes qualitatively apart for the Planck order shifts and macroscopic shifts and the response rate for the microscopic shift is not perturbatively small. In order to identify the shift parameter $\ell_1$ the strategy could be to track the response rate as a  function of parameter $T$ over a range of its value and identifying the appropriate curve thereof, in Fig.~\ref{resrate}, addressing any potential multivaluedness the response rate could have as a function of $\ell_1$.

\begin{figure}[H]
    \centering
    \begin{subfigure}[b]{0.48\textwidth}
    \centering
    \includegraphics[width=\textwidth]{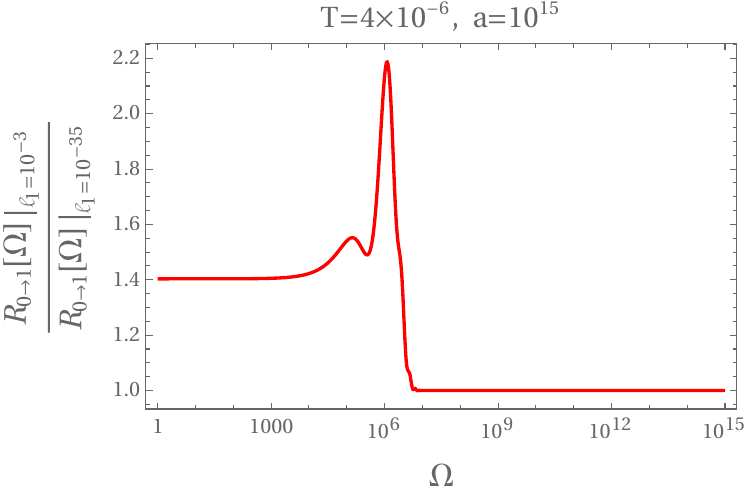}
    \subcaption[]{}
    \label{ratio1}
    \end{subfigure}
    \hfill
    \begin{subfigure}[b]{0.48\textwidth}
    \centering
    \includegraphics[width=\textwidth]{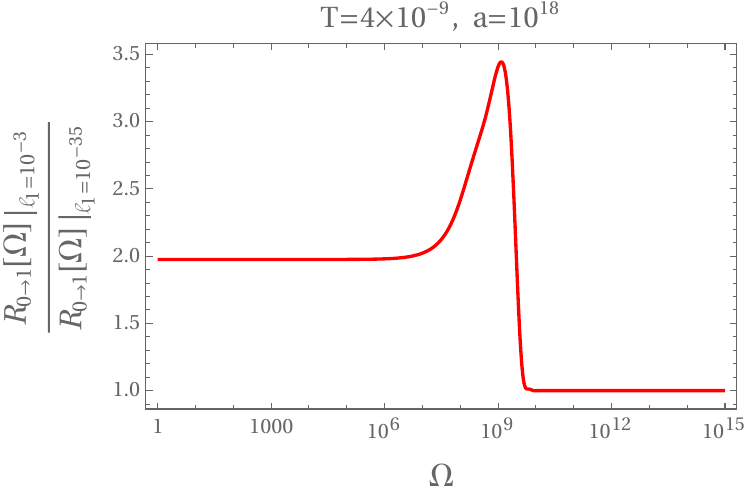}
    \subcaption[]{}
    \label{ratio2}
    \end{subfigure}
    \caption{The ratio of response rate for the macroscopic shift to the response rate for Planck shift as a function of the transition frequency with detectors kept on for $T=4\times10^{-6} s$ and acceleration $a=10^{15}ms^{-2}$ in Fig. \ref{ratio1} and detectors kept on for $T=4\times10^{-9} s$ and with acceleration $a=10^{18}ms^{-2}$ in Fig. \ref{ratio2}.}
    \label{resratio}
\end{figure}

\subsection{Detector Energy Spacing}
In order to find an optimal detector for differentiating micro and macro shifts of horizons, we need to identify the domain of the energy gap for which the response, as well as its difference between macroscopic and microscopic shifts, is appreciable for a given acceleration. As discussed earlier, the different responses for different shifts come from the interplay of thermal contribution and series (inertial-like) contribution in \eqref{tpft}. The inertial-like contribution has the  shift parameter only in the phase and thus the response rate governed by it is independent of the shift, while the thermal term has $(a_1\ell_1)^{2/a_1T}$ as a prefactor. From the analytical expression in \eqref{tpft}, we see the thermal limit for this detector is when it is switched on for a duration longer than the time scales of the system, i.e., when we have $a_1T/c\gg 1$ and $\Omega T\gg 1$. For the parameter regimes that ensure robust numerical analysis in our exploration, the exact thermality is not visible since $a_1T/c\leq 1$. For large $\Omega$, the inertial-like contribution dominates over the thermal part, which is exponentially suppressed, and thus the response is expected to be independent of the shift. On the other hand, we expect the thermal contribution to increase as $\Omega$ decreases, and at some point, it should become comparable to the inertial-like term and hence of relevance for the total response, thereby making the response shift dependent. Thus, for the low energy gap detectors, the difference between micro and macroscopic response gets more pronounced for a given acceleration and switch-on time.

In Fig. \ref{resratio}, we plot the ratio of the response rate for the macroscopic shift to the microscopic Planck scale shift as a function of the detector energy gap $\Omega$. Regardless of the acceleration of the detector, we see that the ratio is qualitatively different from unity for small $\Omega$ and settles to one for large $\Omega$, as expected. As argued, for $\Omega T\gg 1$, the thermal contribution to the response is negligible for both micro and macroscopic shifts, and the inertial-like term determines the response for both cases, giving unit value to the ratio. On the other hand, for $\Omega T\leq 1$, the thermal contribution is of the same order as the inertial-like term for the macroscopic shift. In this limit, the thermal contribution for the microscopic shift will be suppressed by a factor $(\ell_{Pl}/\ell_{macro})^{2/a_1T}$ compared to the macroscopic shift, hence it will become insignificant in total response. Thus, we can see that in the limit $\Omega T\leq 1$, the macroscopic shift response is comprised of both the thermal and inertial-like terms, both roughly of similar order, while the microscopic shift response is essentially determined only by the inertial-like term. Therefore, we can expect the macroscopic shift response to be $1-2$ times larger than the microscopic shift case, as indeed can be seen from Fig. \ref{resratio}.

The ratio further attains a maximum near $\Omega T\sim O(1)$, while it settles to a constant value greater than one for $\Omega T\ll 1$. Thus, we find that the Planck order effects become quickly irrelevant for a large transition frequency detector, i.e., $\Omega T\gg 1$, while the microscopic shifts can be efficiently and distinguishably picked up by detectors with smaller transition frequency with the strongest imprints being in the regime $\Omega T\sim 1$. Therefore, an optimal detector would be one with $a_1 T \sim c$, which translates to the condition $\Omega c/a_1 \sim 1$ with the timescale of operation $1/\Omega$. For such detectors, the contrast between a microscopic shift in the Rindler wedges and a macroscopic shift will be most pronounced.

\section{Discussion}

In this work, we analyze if an Unruh-DeWitt probe on an accelerated trajectory can sense shifts in the Rindler wedges they live in, as compared to that of another Rindler observer. The analysis is motivated by the field theoretic considerations that suggest that the global objects, such as the number expectation, turn on to a full thermality whenever the Rindler wedges do not completely overlap and mismatch even by the tiniest microscopic shift possible -- i.e., the Planck scale \cite{Lochan:2021pio}. However, such computations require the full spacetime geometry to be completely known beforehand. Thus, it is not very clear if there can be realistic detectors operating in finite regions of spacetime which can respond in the above-mentioned fashion. 

In this article, we employ a realistic Unruh-DeWitt detector to analyze the nested system of Rindler frames. We see that the UDW detector, which operates for infinite time responds in a fashion exactly in tune with the number operator expectation. Thus such detectors will respond with a thermal character whenever two accelerated observers do not asymptote to the same null ray in flat space. Though an infinite time detector robustly signals any mismatch in the Rindler wedge of two Rindler observers, it also loses information about the amount of shift. Thus such detectors cannot tell with definiteness if the shift is of Planck scale or macroscopic.  However, it turns out that a finite-time detector on an accelerated trajectory is naturally tuned to capture the tiniest shift in the perceived Rindler horizon compared to any other Rindler wedge. Unlike an infinite time detector, a finite time detector carries a shift dependency in its response that becomes prominent for a detector with energy gap $\Omega c/a \sim 1$ and that operates for timescale $T c/a \sim 1$ and beyond. For such detectors, the response for a macroscopic shift and microscopic shift are clearly robust and differentiable. It is interesting to note that the response for the Planck shift in both the finite and infinite time detector is of a similar order to that of the Unruh thermality. Therefore, the accelerated detector and fields in the Fock space of any Rindler wedge provide a combination well suited for observing Planck scale shifts in the Rindler causal wedges.

It is recently argued that the thermal effect in the UDW detector can be made observationally strong even for moderately small accelerations by employing appropriate boundary conditions \cite{Stargen:2021vtg}. In addition, there are many proposals that try to bring the Unruh effect in the realm of observational verification \cite{Crispino:2007eb}. Thus, if the Unruh thermality can be efficiently captured in any setup, it is likely that even the definite Planck scale signatures can also be put to the test in those setups.

This problem has a potential correspondence with accelerated observers in the black hole exterior geometry. Since the analysis really scrutinizes if two accelerated observers will asymptotically approach the same null ray or not, it can be employed between two exterior observers in the exterior of a black hole geometry in which one of the observers asymptotically approaches the event horizon while the other just crosses over due to the growth of the horizon caused by an in-fall of some microscopic mass. Thus, one needs to generalize these computations to a curvature-full and more realistic $3\oplus1$ dimensional spacetime. Moreover, even the field theoretic computations should be generalized for such finite time operations to gain insights if the emission from the black hole also contains rich information of its horizon shift. In the context of cosmology, similar analysis may potentially address the effects revealing shifts in cosmological horizons due to the expansion of the universe. These analyses will hopefully be performed in subsequent studies.

\section{Acknowledgments}
Research of KL is partially supported by SERB, Govt. of India through a MATRICS research grant no. MTR/2022/000900. HSS would like to acknowledge the financial support from the University Grants Commission, Government of India, in the form of Junior Research Fellowship (UGC-CSIR JRF/Dec-2016/503905) and the Core Research Grant CRG/2021/003053 from the Science and Engineering Research Board, India.

\appendix

\section{Derivation of finite time detector response}\label{App1}
In this section, we derive the expression for the response function for the detector that is switched on for a finite time and check the robustness of the result for different physical scenarios. The integral of interest is Eq. \ref{io1} of the main paper
\begin{align}
	I_1(\omega)=\int_{-\infty}^\infty dk\;\tilde{\chi}(k)\int_{-\infty}^\infty d\tau \;e^{-i\left(\Omega+ \omega\frac{a_1}{a_0}-k\right)\tau}\left(1+a_1\ell_1e^{a_1\tau}\right)^{i\frac{ \omega}{a_0}},\label{io2}
\end{align}
where $\tilde{\chi}(k)$ is the Fourier transform of the window function. For the case of a window function with exponential cutoff $\chi(t)=e^{-|t|/T}$, we have
\begin{align}
	\tilde{\chi}(k)=\frac{T}{\pi(1+k^2T^2)}.
\end{align}
The time integral in Eq. \eqref{io2} can be solved using the result \cite{Oberhettinger1990}
\begin{align}
	\int_{-\infty}^{\infty}dx \;e^{(\epsilon+ip)x}(1+Ae^{x})^{-\epsilon_2-iq} &=\frac{A^{-\epsilon-ip}\Gamma(\epsilon_1+i(q-p))\Gamma(\epsilon+ip)}{\Gamma(\epsilon_2+iq)},\label{ober}
\end{align}
where $p\;\&\;q\;\in\;\mathbb{R}$ and the regulators behave as $0<\epsilon_1<\epsilon_2$ and $\epsilon_2-\epsilon_1=\epsilon>0$. The regulators in this analysis are of particular importance, and by retaining these, we arrive at
\begin{align}
	I_1(\omega)=\frac{T}{\pi}\int_{-\infty}^\infty dk\;\frac{(a_1\ell_1)^{-\epsilon_1+i\left(\frac{\Omega-k}{a_1} +\frac{\omega} {a_0}\right)} \Gamma\left(\epsilon_2-\epsilon_1+i\frac{\Omega-k}{a_1}\right)\Gamma\left(\epsilon_1-i\left( \frac{\Omega-k}{a_1}+\frac{\omega} {a_0}\right)\right)}{a_1(1+k^2T^2)\Gamma\left(\epsilon_2-i\frac{\omega}{a_0} \right)}.
\end{align}
This integration can be solved by using the residue theorem, in which the regulators determine the location of the poles. It is convenient to work with
\begin{align}
	I_1(\omega)&=-\frac{i\;T(a_1\ell_1)^{i\left(\frac{\Omega}{a_1} +\frac{\omega} {a_0}\right)}}{\pi\Gamma\left(\epsilon_2-i\frac{\omega}{a_0} \right)}\int_{-i\infty}^{i\infty} dx(a_1\ell_1)^{-x}\frac{1}{1-a_1^2x^2T^2}\Gamma\left(\epsilon_2-\epsilon_1+i\frac{\Omega}{a_1}-x\right)\Gamma\left(\epsilon_1-i\left( \frac{\Omega}{a_1}+\frac{\omega} {a_0}\right)+x\right)\\
	&=-\frac{i\;T(a_1\ell_1)^{i\left(\frac{\Omega}{a_1} +\frac{\omega} {a_0}\right)}}{\pi\Gamma\left(\epsilon_2-i\frac{\omega}{a_0} \right)}\int_{-i\infty}^{i\infty} dx(a_1\ell_1)^{-x}\frac{1}{1-a_1^2x^2T^2}\Gamma\left(\beta-x\right)\Gamma\left(\alpha+x\right).
\end{align}
The direction of the closure of the contour is determined by whether $a_1\ell_1>1$ or $a_1\ell_1<1$ in order to satisfy Jordan's lemma. In the first case, the contour is closed from $Re(x)>0$, and in the second case, from $Re(x)<0$. In this analysis, we are interested in the shifts of extremely small magnitude; therefore, the second case is preferred. We can arrive at the expression for the first case following the same derivation. The poles of the integrand are at
\begin{align}
	\Gamma(\alpha+x)&:\qquad \alpha+\alpha_n=-n\implies \alpha_n=-n-\alpha;\qquad\forall\; n\;\in \mathbb{N}_0,\\
	\Gamma(\beta-x)&:\qquad \beta-\beta_n=-n\implies \beta_n=n+\beta,\\
	\frac{1}{1-a_1^2x^2T^2}&:\qquad x_\pm=\pm\frac{1}{a_1T}.
\end{align}
Residues at respective poles are
\begin{align}
	\alpha_n:&\qquad Res_{\alpha_n}=\frac{(-1)^n(a_1\ell_1)^{\alpha+n}\Gamma(\alpha+\beta+n)}{(1-a_1^2T^2(\alpha+n)^2)n!},\\
	\beta_n:&\qquad Res_{\beta_n}=\frac{(-1)^{n+1}(a_1\ell_1)^{-\beta-n}\Gamma(\alpha+\beta+n)}{(1-a_1^2T^2(\beta+n)^2)n!},\\
	x_\pm:&\qquad Res_{x_\pm}=\mp(a_1\ell_1)^{\mp\frac{1}{a_1T}} \frac{\Gamma\left(\beta\mp\frac{1}{a_1T}\right)\Gamma\left(\alpha\pm\frac{1}{a_1T}\right)}{2a_1T}.
\end{align}
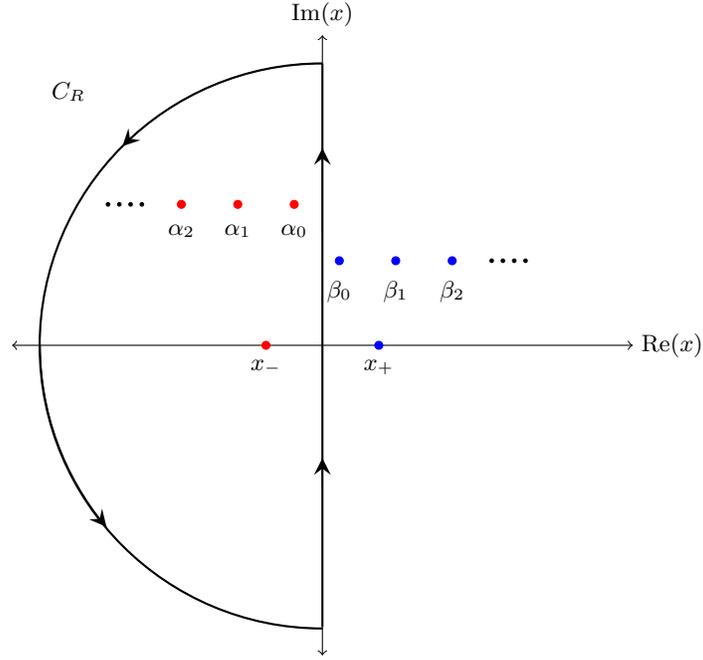
\begin{figure}[H]
	\centering
	\begin{tikzpicture}[scale=0.75]
		\draw[help lines, color=gray!30, dashed] (-4.9,-4.9) ;
		\draw[<->] (-5.5,0)--(5.5,0) node[right]{$\text{Re}(x)$};
		\draw[<-] (0,-5.5)--(0,-5);
		\draw[->] (0,5)--(0,5.5) node[above]{$\text{Im}(x)$};
		\draw [thick] (0,-5)--(0,5);
		\filldraw [red] (-0.5,2.5) circle (2pt);
		\draw (-0.5,2.3) node[below]{$\alpha_0$};
		\draw (-1.5,2.3) node[below]{$\alpha_1$};
		\draw (-2.5,2.3) node[below]{$\alpha_2$};
		\draw (-1,-0.1) node[below]{$x_-$};
		\draw [decoration={markings,mark=at position 1 with
			{\arrow[scale=2,>=stealth]{>}}},postaction={decorate}] (0,-2.5)--(0,-2);
		\draw [decoration={markings,mark=at position 1 with
			{\arrow[scale=2,>=stealth]{>}}},postaction={decorate}] (0,2)--(0,3.5);
		\draw [decoration={markings,mark=at position 1 with
			{\arrow[scale=2,>=stealth]{>}}},postaction={decorate}] (-5,0) arc (180:220:142.6pt);
		\draw [decoration={markings,mark=at position 1 with
			{\arrow[scale=2,>=stealth]{>}}},postaction={decorate}] (0,5) arc (90:135:142.6pt);
		\filldraw [red] (-1,0) circle (2pt);
		\filldraw [red] (-1.5,2.5) circle (2pt);
		\filldraw [red] (-2.5,2.5) circle (2pt);
		\filldraw [black] (-3.8,2.5) circle (0.9pt);
		\filldraw [black] (-3.2,2.5) circle (0.9pt);
		\filldraw [black] (-3.4,2.5) circle (0.9pt);
		\filldraw [black] (-3.6,2.5) circle (0.9pt);
		\filldraw [blue] (0.3,1.5) circle (2pt);
		\filldraw [blue] (1.3,1.5) circle (2pt);
		\filldraw [blue] (2.3,1.5) circle (2pt);
		\filldraw [blue] (1,0) circle (2pt);
		\draw (1,-0.1) node[below]{$x_+$};
		\draw (0.3,1.3) node[below]{$\beta_0$};
		\draw (1.3,1.3) node[below]{$\beta_1$};
		\draw (2.3,1.3) node[below]{$\beta_2$};
		\filldraw [black] (3.0,1.5) circle (0.9pt);
		\filldraw [black] (3.2,1.5) circle (0.9pt);
		\filldraw [black] (3.4,1.5) circle (0.9pt);
		\filldraw [black] (3.6,1.5) circle (0.9pt);
		\draw (-4.5,4.5) node{$C_R$};
		\draw [thick] (0,5) arc (90:270:142.6pt);
	\end{tikzpicture}
	\caption{Contour relevant for the case when $a_1\ell_1<1$. The residues of poles at red points contribute in this case, whereas the residues of poles at blue points contribute in the case of $a_1\ell_1>1$}
	\label{Con1}
\end{figure}

In the case of $a_1\ell_1<1$, the relevant contour is a closed loop consisting of the imaginary axis and the semicircular arc of infinite radius on the left side, $C_\infty\cup\mathbb{R}$. The poles that are enclosed in the contour are $\alpha_n$ and $x_-$ as shown in Fig. \ref{Con1}, and the contour integral is given by
\begin{align}
	\oint dx(a_1\ell_1)^{-x}\frac{1}{1-a_1^2x^2T^2}\Gamma\left(\beta-x\right)\Gamma\left(\alpha+x\right)=2\pi i\left(\sum_{n=0}^\infty Res_{\alpha_n}+Res_{x_-}\right).
\end{align}
The integral along the semicircular arc $C_\infty$ vanishes following Jordan's lemma, and therefore the integral of interest takes the form
\begin{align}
	I_1(\omega)=\frac{2T(a_1\ell_1)^{i\left(\frac{\Omega}{a_1} +\frac{\omega} {a_0}\right)}}{\Gamma\left(\epsilon_2-i\frac{\omega}{a_0} \right)}\left[\sum_{n=0}^\infty\left(\frac{(-1)^{n}(a_1\ell_1)^{\alpha+n}\Gamma(\alpha+\beta+n)}{(1-a_1^2T^2(\alpha+n)^2)n!}\right)+\frac{(a_1\ell_1)^{\frac{1}{a_1T}}\Gamma\left(\beta+\frac{1}{a_1T}\right)\Gamma\left(\alpha-\frac{1}{a_1T}\right)}{2a_1T}\right].\label{AIo}
\end{align}
Since $|I_1(\omega)|^2$ appears in the expression of response function, we omitted the outside phase factor in the main text. At this point, we can take the limit $\epsilon_{1/2}\rightarrow 0$ in the expressions. The first thing we will check is the convergence of infinite series.
\begin{align}
	\lim_{n\rightarrow\infty}\left|\frac{Res_{\alpha_{n+1}}}{Res_{\alpha_n}}\right|&=\lim_{n\rightarrow\infty}\left|\frac{(-1)^{n+1}(a_1\ell_1)^{\alpha+n+1}\Gamma(\alpha+\beta+n+1)}{(1-a_1^2T^2(\alpha+n+1)^2)(n+1)!}\frac{(1-a_1^2T^2(\alpha+n)^2)n!}{(-1)^{n}(a_1\ell_1)^{\alpha+n}\Gamma(\alpha+\beta+n)}\right|\nonumber\\
	&=\lim_{n\rightarrow\infty}\left|-\frac{a_1\ell_1(\alpha+\beta+n)(1-a_1^2T^2(\alpha+n)^2)}{(1-a_1^2T^2(\alpha+n+1)^2)(n+1)}\right|=|a_1\ell_1|<1.
\end{align}
Therefore, the series converges according to Cauchy's ratio test. Next, we check the consistency of the result obtained looking at the behavior of the response rate under different physical limits. 

\subsection{Eternal and instantaneous switching limit}\label{App11}
In the $T\rightarrow\infty$ limit, the window function becomes unity, representing the case of eternal switching, and one expects to recover the infinite-time result. For this limit, each term in the infinite series in Eq. \eqref{AIo} behaves as $T^{-1}$ and thus vanishes. The last term in Eq. \eqref{AIo} is
\begin{align}
	\lim_{T\rightarrow\infty}I_1(\omega)=\lim_{T\rightarrow\infty}\frac{(a_1\ell_1)^{\frac{1}{a_1T}}\Gamma\left(i\frac{\Omega}{a_1}+\frac{1}{a_1T}\right)\Gamma\left(-i\left(\frac{\Omega}{a_1}+\frac{\omega}{a_0}\right)-\frac{1}{a_1T}\right)}{a_1\Gamma\left(-i\frac{\omega}{a_0}\right)}=\frac{\Gamma\left(i\frac{\Omega}{a_1}\right)\Gamma\left(-i\left(\frac{\Omega}{a_1}+\frac{\omega}{a_0}\right)\right)}{a_1\Gamma\left(-i\frac{\omega}{a_0}\right)}
\end{align}
which will give us the thermal response. At the other asymptotic limit $T\rightarrow 0$, the window function is sharply peaked at the origin, and we expect the response function to vanish. Again, the infinite series is proportional to $T$ in this limit and therefore vanishes. The relevant asymptotic expressions are
\begin{align}
	\left|\Gamma\left(i\frac{\Omega}{a_1}+\frac{1}{a_1T}\right)\Gamma\left(-i\left(\frac{\Omega}{a_1}+\frac{\omega}{a_0}\right)-\frac{1}{a_1T}\right)\right|^2\xrightarrow{T\rightarrow 0}\left|\pi a_1T\csc\left[\pi  \left(-\frac{1}{a_1T}-i \left(\frac{\omega}{a_0} +\frac{\Omega}{a_1} \right)\right)\right]\right|^2.
\end{align}
For the last term in Eq. \eqref{AIo}, we have
\begin{align}
	\lim_{T\rightarrow 0}|I(\omega)|^2&=\lim_{T\rightarrow 0}\left|\frac{(a_1\ell_1)^{\frac{1}{a_1T}}\Gamma\left(i\frac{\Omega}{a_1}+\frac{1}{a_1T}\right)\Gamma\left(-i\left(\frac{\Omega}{a_1}+\frac{\omega}{a_0}\right)-\frac{1}{a_1T}\right)}{a_1\Gamma\left(-i\frac{\omega}{a_0}\right)}\right|^2\nonumber\\
	&=\lim_{T\rightarrow 0}\left|\frac{\pi T(a_1\ell_1)^{\frac{1}{a_1T}}\csc\left[\pi  \left(-\frac{1}{a_1T}-i \left(\frac{\omega}{a_0} +\frac{\Omega}{a_1} \right)\right)\right]}{\Gamma\left(-i\frac{\omega}{a_0}\right)}\right|^2
\end{align}
Since $a_1\ell_1<1$, the term $(a_1\ell_1)^{1/T}$ is decaying exponentially whereas the $\csc$ term is highly oscillating but with finite amplitude as $T\rightarrow 0$. Therefore, the response function and thus the transition probability vanishes in this case.

\subsection{Vanishing shift limit}\label{App12}
In the limit of the vanishing shift, the thermal equivalence is lost, and we expect the inertial response in this case. With the window function under consideration $\chi=e^{-|t|/T}$, the response of a detector on the inertial trajectory for the Minkowski vacuum is given by
\begin{align}
	P_{0\rightarrow1}=&\bar{m}_{10}\int_{-\infty}^\infty dt\int_{-\infty}^\infty dt'e^{-i\Omega(t-t')}\chi(t)\chi(t')\braket{0_M|\phi(x)\phi(x')|0_M}\nonumber\\
    =&\frac{\bar{m}_{10}^2}{2\pi}\int^\infty_0\frac{d\omega}{\omega}\frac{4T^2}{(1+(\Omega+\omega(1-v)\gamma)^2T^2)^2}.
\end{align}
where $v$ is the velocity of the detector, and $\gamma$ is the Lorentz factor. On the other hand, all terms in Eq. \eqref{AIo} vanish except for the leading-order term in infinite series in the limit of $\ell_1\rightarrow 0$.
\begin{align}
	I_1(\omega)=\lim_{\ell_1\rightarrow 0}\frac{2T(a_1\ell_1)^{i\left(\frac{\Omega}{a_1} +\frac{\omega} {a_0}\right)}}{\Gamma\left(\epsilon_2-i\frac{\omega}{a_0} \right)}\left(\frac{(a_1\ell_1)^{\alpha}\Gamma(\alpha+\beta)}{(1-a_1^2T^2\alpha^2)}\right)=\frac{2T}{1+a_1^2T^2\left(\frac{\Omega}{a_1}+\frac{\omega}{a_0}\right)^2},
\end{align}
where $\alpha=-i(\Omega/a_1+\omega/a_0)$ and $\beta=i\Omega/a_1$, leading to the response function
\begin{align}
	P_{0\rightarrow1}=\bar{m}_{10}^2\int^\infty_0\frac{d\omega}{2\pi\omega}\frac{4T^2}{\left(1+T^2\left(\Omega+\frac{a_1\omega}{a_0}\right)^2\right)^2}.\label{InRes}
\end{align}
The response function, in this case, has qualitative characteristics of the inertial case, with the ratio of accelerations of different Rindler frames $a_1/a_0$ playing the role of rapidity defined as $e^{-w}=(1-v)\gamma=\sqrt{(1-v)/(1+v)}$. Since the system under consideration is $1\oplus 1$ dimensional scalar field, the response function has IR divergence, related to the IR ambiguity in the Wightman function of a scalar field in $1\oplus 1$ dimensions \cite{Louko:2014aba}. This integral can be regularized by separating it into a divergent and finite part by introducing an IR cutoff \cite{Svaiter:1992xt,Estrada}
\begin{align}
    \int^\infty_\epsilon\frac{d\omega}{2\pi\omega}\frac{4T^2}{\left(1+T^2\left(\Omega+a\omega\right)^2\right)^2}=&-\frac{2 \log (a T \epsilon )}{\pi  \left(T^2 \Omega ^2+1\right)^2}+\frac{T^2 }{2 \pi  \left(T^2 \Omega ^2+1\right)^2}\bigg(-T \Omega  \left(\pi  T^2 \Omega ^2-2 T \Omega +3 \pi \right)\nonumber\\
    &+2 \log \left(T^2 \Omega ^2+1\right)+2 T \Omega  \left(T^2 \Omega ^2+3\right) \tan ^{-1}(T \Omega )-2\bigg)+O(\epsilon),
\end{align}
or using dimensional regularization-motivated expression
\begin{align}
    \int^\infty_0\frac{d\omega}{2\pi\omega^{1-\epsilon}}\frac{4T^2}{\left(1+T^2\left(\Omega+a\omega\right)^2\right)^2}=\frac{T^2 }{2 \pi  \left(T^2 \Omega ^2+1\right)^2}\bigg[&\frac{4}{\epsilon}+2 T^2 \Omega ^2-2+\Re\big\{\left(2+i T \Omega  \left(T^2 \Omega ^2+3\right)\right)\nonumber\\
    &\times\log (T \Omega +i)\big\}\bigg]+O(\epsilon)
\end{align}
or using Hadamard form of the Wightman function of the massless scalar field in $1\oplus1$ dimension \cite{Louko:2014aba}
\begin{align}
    P_{0\rightarrow1}=-\frac{\bar{m}_{10}^2}{2\pi}\int_{-\infty}^\infty d\tau\int_{-\infty}^\infty d\tau'e^{-i\Omega(\tau-\tau')}\chi(\tau)\chi(\tau')\log[\mathrm{m}_0(\epsilon+i(\tau-\tau'))].
\end{align}
Where $\mathrm{m}_0$ is a positive constant of dimension inverse length understood as IR cutoff leading to the response function
\begin{align}
    P_{0\rightarrow1}=-\frac{2 T^2 \log (\mathrm{m}_0T)}{\pi  \left(T^2 \Omega ^2+1\right)^2}+\frac{T^2 }{2 \pi  \left(T^2 \Omega ^2+1\right)^2}\bigg(&2 \log \left(T^2 \Omega ^2+1\right)-2-T \Omega  \left(\pi  T^2 \Omega ^2-2 T \Omega +3 \pi \right)\nonumber\\
    &+2 T \Omega  \left(T^2 \Omega ^2+3\right) \tan ^{-1}(T \Omega )\bigg)
\end{align}
\begin{figure}
	\centering
            \begin{subfigure}[b]{0.44\textwidth}
    \centering
    \includegraphics[width=\textwidth]{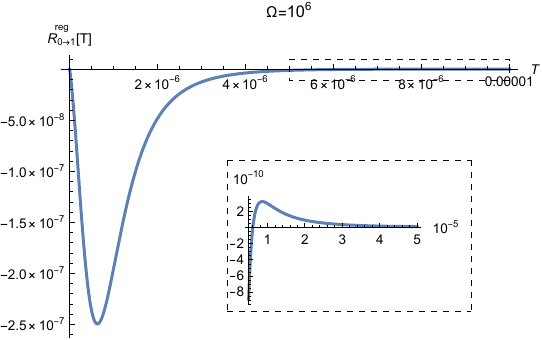}
    \subcaption[]{}
    \label{f1}
    \end{subfigure}\\
    \hfill
    \begin{subfigure}[b]{0.44\textwidth}
    \centering
    \includegraphics[width=\textwidth]{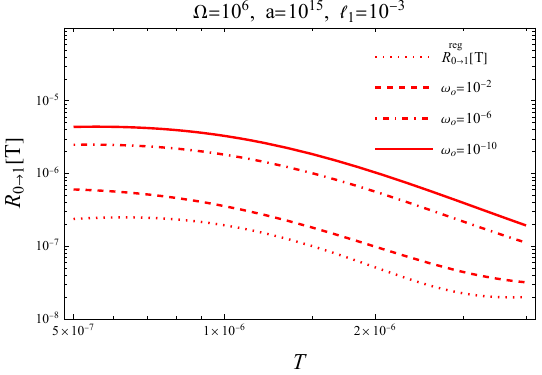}
    \subcaption[]{}
    \label{f2}
    \end{subfigure}
    \hfill
    \begin{subfigure}[b]{0.44\textwidth}
    \centering
    \includegraphics[width=\textwidth]{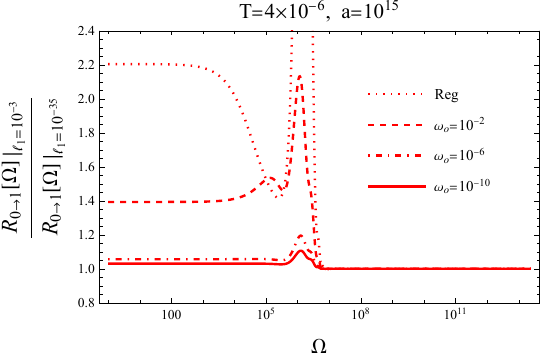}
    \subcaption[]{}
    \label{f3}
    \end{subfigure}
		\caption{The behavior of regularized inertial response as a function of switch-on time for fixed energy gap in Fig. \ref{f1}. Comparison of the behavior of response rate for different IR cutoffs and regularization in Figs. \ref{f2} and \ref{f3}. In the second frame, the response rate as a function of detector switch-on time with fixed energy gap and acceleration for macroscopic shift and the ratio of detector response for macroscopic shift to microscopic shift as a function of the energy gap with fixed acceleration and detector switch-on time in the third frame. The qualitative features of the response rate are independent of the cutoff chosen. We have used $\bar{m}_{10}^2/\hbar^2=10^6$ in these plots.}
		\label{diffcutoff}
\end{figure}
Although with different origins, the regularized expressions corresponding to the three schemes have the same behavior shown in Fig. \ref{f1}, with expressions matching exactly for the first and third cases. In general, the regularized contribution starts from the origin, decreases to attain a global minimum at $T\sim0.6/\Omega$, and starts increasing to cross the x-axis at $T\sim6.1/\Omega$, attaining a global maximum at $T\sim8.6/\Omega$. As discussed in the main text, the regime of importance in our analysis is $\Omega T\sim 1$; both features of regularized contribution will play a role. To facilitate the comparison of the response rate with the IR cut-off and of the response rate with regularized contribution, we consider the absolute value of the regularized contribution. In our analysis, we work with the arbitrary IR cutoff to deal with the divergence, and we have shown that the central results of our analysis are independent of the chosen cutoff or regularized expression, as shown in Figs. \ref{f2} and \ref{f3}.

\section{Convergence of the finite time response}\label{App2}
In this section, we discuss the convergence of the integral appearing in the finite time response in Eq. \ref{tpft} of the main paper. First, the inertial response has IR divergence as seen in the integral in Eq. \eqref{InRes}, which is related to the IR ambiguity in the Wightman function of the scalar field in $1\oplus 1$ dimension \cite{Louko:2014aba}. To numerically integrate, we are introducing an IR cutoff that leads to a finite inertial response. For the thermal part of the response, the integral of interest is
\begin{align}
	I(T,\Omega,a_1)=\left| \Gamma\left(i\frac{\Omega}{a_1}+\frac{1}{a_1T}\right)\right|^2\int_0^\infty\frac{d\tilde{\omega}}{2\pi\tilde{\omega}}\left|\frac{\Gamma\left(-i\left(\frac{\Omega}{a_1}+\tilde{\omega}\right)-\frac{1}{a_1T}\right)}{a_1\Gamma\left(-i\tilde{\omega}\right)}\right|^2,
\end{align}
where we have rescaled the integration variable $\omega$. The asymptotic behavior of the integrand at the limits of integration is
\begin{align}
	\lim_{\tilde{\omega}\rightarrow 0}\frac{1}{\tilde{\omega}}\left|\frac{\Gamma\left(-i\left(\frac{\Omega}{a_1}+\tilde{\omega}\right)-\frac{1}{a_1T}\right)}{\Gamma\left(-i\tilde{\omega}\right)}\right|^2&=\left|\Gamma\left(-i\frac{\Omega}{a_1}-\frac{1}{a_1T}\right)\right|^2\tilde{\omega}(1+O(\tilde{\omega})),\label{o0}
\end{align}
\begin{align}
	\lim_{\tilde{\omega}\rightarrow\infty}\frac{1}{\tilde{\omega}}\left|\frac{\Gamma\left(-i\left(\frac{\Omega}{a_1}+\tilde{\omega}\right)-\frac{1}{a_1T}\right)}{\Gamma\left(-i\tilde{\omega}\right)}\right|^2&=\tilde{\omega}^{-1-1/2a_1T}\left(e^{-\pi\Omega/a_1}+O(\tilde{\omega}^{-1})\right).\label{oin}
\end{align}
For the case at hand, the integrand has the appropriate behavior at both limits for the integral to converge. The integral is logarithmically divergent at the limit of $T\rightarrow\infty$, as expected \cite{Lochan:2021pio}. Although the falloff of the tail of the integrand for finite $T$ and large $\omega$ is faster than $\omega^{-1}$, one still has to be careful when numerically integrating it.

Ideally, for an improper integral with the upper limit as infinite, one needs to identify a cutoff beyond which the contribution of the area under the tail can be neglected. For the behavior of the integrand under consideration, such a universal cutoff $\omega_o$ does not exist. In fact, the cutoff in this case is highly sensitive to $a_1$ and $T$, as these parameters appear in the exponent in Eq. \eqref{oin}. The strategy we will employ here is, to identify the window of the parameter $T$ for a particular acceleration, for which we can trust the numerical results with an appropriate cutoff. 

\section{Ambiguity in defining finite time response}\label{App3}
\begin{figure}[H]
	\centering
		\includegraphics[width=0.45\textwidth]{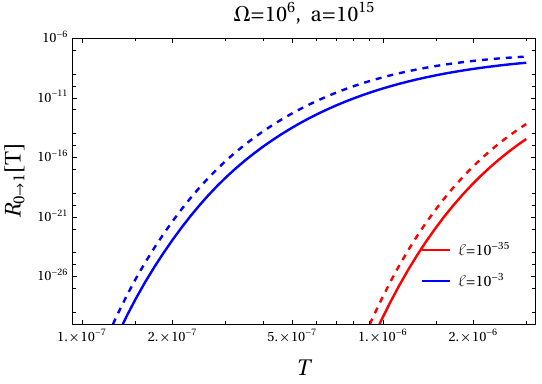}
        \includegraphics[width=0.48\textwidth]{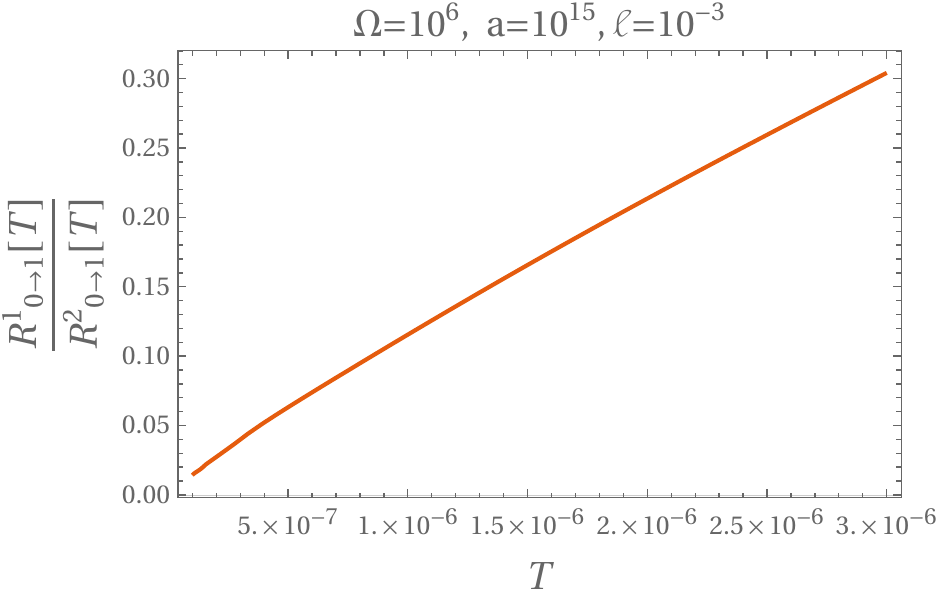}
        \includegraphics[width=0.48\textwidth]{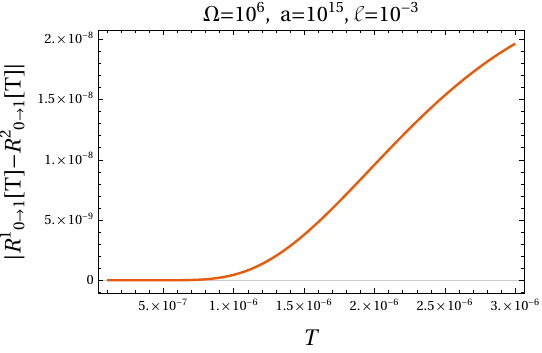}
		\caption{The thermal contribution to the response rate as a function of the width of the window function for the shifts of Planck length (red curves) and macroscopic order (blue curves). Here, the solid profiles correspond to the response rate defined in Eq. \eqref{RR1}, and the dashed profiles correspond to the response rate defined in Eq. \eqref{RR2}.}
		\label{therresrate}
\end{figure}

The response rate can be defined by either dividing the response function by the time the detector is switched on or by taking the time derivative of the response function. In the limit, $T\rightarrow\infty$, these two notions are expected to agree. In this appendix, we estimate the difference in two prescriptions for the finite $T$ case. For simplicity, we consider the case of thermal contribution to the response only and restoring the factors of $c$ and $\hbar$, the two notions of the response rate are given by

\begin{align}
	R^1_{0\rightarrow1}(\Omega,a_1,T)&=\frac{1}{T}\left(\frac{a_1\ell_1}{c^2}\right)^{\frac{2c}{a_1T}}\int_0^\infty\frac{d\omega}{2\pi\omega}\left|\frac{c\Gamma\left(i\frac{\Omega c}{a_1}+\frac{c}{a_1T}\right)\Gamma\left(-i\left(\frac{\Omega c}{a_1}+\frac{\omega c} {a_0}\right)-\frac{c}{a_1T}\right)}{a_1\Gamma\left(-i\frac{\omega c}{a_0}\right)}\right|^2,\label{RR1}\\
	R^2_{0\rightarrow1}(\Omega,a_1,T)&=\frac{d}{dT}\left[\left(\frac{a_1\ell_1}{c^2}\right)^{\frac{2c}{a_1T}}\int_0^\infty\frac{d\omega}{2\pi\omega}\left|\frac{c\Gamma\left(i\frac{\Omega c}{a_1}+\frac{c}{a_1T}\right)\Gamma\left(-i\left(\frac{\Omega c}{a_1}+\frac{\omega c} {a_0}\right)-\frac{c}{a_1T}\right)}{a_1\Gamma\left(-i\frac{\omega c}{a_0}\right)}\right|^2\right].\label{RR2}
\end{align}
We have plotted the response rate as a function of the width of the window function and the transition frequency in Fig. \ref{therresrate}. For both definitions of response rate, the functional dependence is the same, although the magnitude is larger for the case of \eqref{RR2}.

%


\end{document}